\documentclass[preprint,prb,aps,superscriptaddress,showpacs]{revtex4-1}
\usepackage{graphicx}
\usepackage{amssymb}
\usepackage{textcomp}
\usepackage{amsmath}
\usepackage[normalem]{ulem} 
\usepackage{soul} 
\def\be{\begin{equation}}
\def\ee{\end{equation}}
\def\bea{\begin{eqnarray}}
\def\eea{\end{eqnarray}}
\def\bdm{\begin{displaymath}}
\def\edm{\end{displaymath}}
\def\ba{\begin{array}}
\def\ea{\end{array}}

\begin{document}

\title{Robust pinning of magnetic moments in pyrochlore iridates}

\author{W.~C. Yang}
\author{W.~K. Zhu}
\altaffiliation[Present address: ]{High Magnetic Field Laboratory, Chinese Academy of Sciences, Hefei 230031, Anhui, China}
\affiliation{Department of Physics, Indiana University, Bloomington IN 47405, USA}
\author{H.~D. Zhou}
\affiliation{Department of Physics and Astronomy, University of Tennessee, Knoxville TN 37996, USA}
\author{L. Ling}
\affiliation{High Magnetic Field Laboratory, Chinese Academy of Sciences, Hefei 230031, Anhui, China}
\author{E.~S. Choi}
\author{M. Lee}
\affiliation{National High Magnetic Field Laboratory, Florida State University, Tallahassee, FL 32310-3706, USA}
\author{Y. Losovyj}
\affiliation{Department of Chemistry, Indiana University, Bloomington IN 47405, USA}
\author{Chi-Ken Lu}
\email{Lu49@ntnu.edu.tw}
\affiliation{Physics Department, National Taiwan Normal University, Taipei 11677, Taiwan}
\author{S.~X. Zhang}
\email[Corresponding author: ]{sxzhang@indiana.edu}
\affiliation{Department of Physics, Indiana University, Bloomington IN 47405, USA}

\date\today
\begin{abstract}

Pyrochlore iridates A$_2$Ir$_2$O$_7$ (A = rare earth elements, Y or Bi) hold great promise for realizing novel electronic and magnetic states owing to the interplay of spin-orbit coupling, electron correlation and geometrical frustration. A prominent example is the formation of all-in/all-out (AIAO) antiferromagnetic order in the Ir$^{4+}$ sublattice that comprises of corner-sharing tetrahedra. Here we report on an unusual magnetic phenomenon, namely a cooling-field induced shift of magnetic hysteresis loop along magnetization axis, and its possible origin in pyrochlore iridates with non-magnetic Ir defects (e.g. Ir$^{3+}$). In a simple model, we attribute the magnetic hysteresis loop to the formation of ferromagnetic droplets in the AIAO antiferromagnetic background. The weak ferromagnetism originates from canted antiferromagnetic order of the Ir$^{4+}$ moments surrounding each non-magnetic Ir defect. The shift of hysteresis loop can be understood quantitatively based on an exchange-bias like effect in which the moments at the shell of the FM droplets are pinned by the AIAO AFM background via mainly the Heisenberg ($J$) and Dzyaloshinsky-Moriya ($D$) interactions. The magnetic pinning is stable and robust against the sweeping cycle and sweeping field up to 35 T, which is possibly related to the magnetic octupolar nature of the AIAO order.

\end{abstract}

\maketitle
    
\section{Introduction}

Spin-orbit coupling (SOC), a relativistic interaction between the spin and the orbit of an electron, is a key ingredient for topologically non-trivial states in condensed matter. ~\cite{TIRev1,TIRev2} Adding strong electron interaction into a spin-orbit coupled system is believed to foster novel magnetic and topological phases,~\cite{Rev1,Rev2, Rev3} such as chiral spin liquid, Weyl semimetals,~\cite{Wan1} topological Mott insulators~\cite{Pesin1} and topological crystalline insulators.~\cite{Fiete1} Among various transition-metal oxides, iridates represent such a unique system with electron correlation and SOC of comparable energy scales.~\cite{Cao1,ArimaScience,Jackeli1,Wan1,Pesin1,Cao2,Cao3,SOC1} In particular, the pyrochlore compounds A$_2$Ir$_2$O$_7$ (A = rare earth elements, Bi or Y) are theoretically predicted to host topological Weyl semimetal (WSM) phases that are characterized by a linear energy dispersion in bulk and open Fermi arcs on surface.~\cite{Wan1} While photoemission studies of these electronic characteristics remain elusive,~\cite{Nakayama} optical conductivity measurements have revealed signatures of WSMs in Rh-doped Nd$_2$Ir$_2$O$_7$~\cite{RhDoped} and undoped Eu$_2$Ir$_2$O$_7$.~\cite{Optical1}

The Weyl semimetal states in pyrochlore iridates are predicted to be accompanied with a non-collinear antiferromagnetic (AFM) order in the Ir$^{4+}$ sublattice that comprises of corner-sharing tetrahedra.~\cite{Wan1,YBKim2} The four Ir$^{4+}$ moments at the vertices of each tetrahedron point either into or outward from its center. This peculiar all-in/all-out (AIAO) magnetic order results from the competition between Heisenberg interaction ($J$), Dzyaloshinsky-Moriya (DM) interaction ($D$), and single-ion anisotropy: with only Heisenberg antiferromagnetic interaction, the Ir$^{4+}$ moments are geometrically frustrated; the magnetic frustration is removed by the DM interaction and the single-ion anisotropy that are enhanced by SOC.~\cite{Wan1,YBKim2} The AIAO AFM order of Ir moments has been experimentally evidenced by a variety of magnetic probe techniques, including resonant X-ray scattering,~\cite{Xray1,Xray2,SmIr} muon spin relaxation studies,~\cite{muSR1,muSR2} and neutron diffraction.~\cite{NdIr}

A prominent feature of the AIAO order is that the four moments in a tetrahedron can be treated as a magnetic octupole whose susceptibility is a third-rank tensor.~\cite{Arima,NatPhysFu} Besides being a weak coupling to magnetic field, the order is unique in the sense that an out-of-plane magnetization can be induced by an in-plane magnetic field.~\cite{NatPhysFu} Another notable characteristic is the existence of two interchangeable magnetic configurations that are linked by a time-reversal operation.~\cite{PRX1} Nontrivial metallic interface states are predicted to exist on the wall of these two time-reversal-related magnetic domains.~\cite{PRX1} Magneto-transport studies have provided strong evidences for the metallic domain walls in Nd$_2$Ir$_2$O$_7$.~\cite{Nd1,Nd2} Such domain walls of low sheet resistance have been visualized in spatially-resolved microwave impedance microscopy measurements.~\cite{Science2015} 

In spite of remarkable electronic and magnetic properties, the synthesis of phase-pure, stoichiometric pyrochlore iridates is often challenging. Indeed, minor impurity phases such as A$_2$O$_3$, Ir and IrO$_2$ were often detected in polycrystalline samples that were grown by conventional solid-state reaction method.~\cite{exp_method1,muSR1,Iridate227Zhu,NdIr,JPCM,NdIr2,NdIr3,anisotropy} Those impurities are either paramagnetic or diamagnetic and hence do not contribute significantly to the observed magnetic phenomena; nevertheless, their presence indicates possible non-stoichiometry of the samples as a result of incomplete reaction in the synthesis process.~\cite{LnIr} Non-stoichiometry has also been reported in single crystals\cite{crystal1,Xray2} and in epitaxial thin films.~\cite{thinfilm1} A remarkable consequence of the non-stoichiometry is the deviation of oxidation states of the elements from their nominal values,~\cite{Iridate227Zhu,IrRu, YIr,thinfilm1} which could have important impact on the magnetic properties of the compounds. In particular, the Ir$^{4+}$ (5d$^5$) has an electronic configuration of $J_{\rm eff} = 1/2$ with a magnetic moment of 1 $\mu_B$ in the atomic limit.~\cite{Cao1} Other oxidation states such as Ir$^{3+}$ (5d$^6$) is nonmagnetic because the $t_{2g}$ states (or $J_{\rm eff} = 1/2$ and 3/2) are completely occupied. In this work, we report on an unusual phenomenon, namely a cooling-field induced shift of magnetic hysteresis loops along the magnetization axis, which may be related to the existence of nonmagnetic defects (e.g. Ir$^{3+}$) in the networks of Ir$^{4+}$ tetrahedra. The shift of magnetic hysteresis loop is robust against the sweeping cycle and the sweeping field up to 35 T. We propose a simple exchange-bias like model in the framework of the AIAO magnetic order to understand the origin of this unusual magnetic behavior. We attribute the robust exchange bias in the magnetization loop to the magnetic octupolar nature of the AIAO order. 

\section{EXPERIMENTAL METHODS}
Due to the difficulty in synthesizing single crystal Lu$_2$Ir$_2$O$_7$, we focused on polycrystalline sample which was prepared by a standard solid-state reaction method, similar to our earlier reports.~\cite{Iridate227Zhu,IridateDPZhu,IridateDPCuIr} High purity Lu$_2$O$_3$ (99.99\%) and IrO$_2$ (99.99\%) powder was mixed at a stoichiometric ratio and was heated in air at 900 $^{\circ}$C and then 1000 $^{\circ}$C for about four days, with intermediate grinding. The growth condition of Y$_2$Ir$_2$O$_7$ sample was reported earlier.~\cite{Iridate227Zhu} X-ray powder diffraction (XRD) measurements were performed using a PANalytical EMPYREAN diffractometer (Cu K $\alpha$ radiation). Both samples are composed of a major pyrochlore phase along with some minor impurities of IrO$_2$ and Lu$_2$O$_3$ (Y$_2$O$_3$) due to incomplete reaction and Ir metal due to decomposition of IrO$_2$ [Supplemental Material Figure S1 and Ref.~\onlinecite{Iridate227Zhu}]. Low field magnetization measurements were carried out in Quantum Design Magnetic Property Measurement Systems (MPMS). High field measurement up to 35 T was carried out in the National High Magnetic Field Laboratory in Tallahassee. X-ray photoelectron spectroscopy (XPS) measurements were conducted in a PHI Versa Probe II system. The XPS spectra were fitted using a standard software package CasaXPS provided by Casa Software, Ltd.

\section{RESULTS AND DISCUSSION}

\begin{figure}
   	\input{epsf}
   	\includegraphics[width=0.8\textwidth]{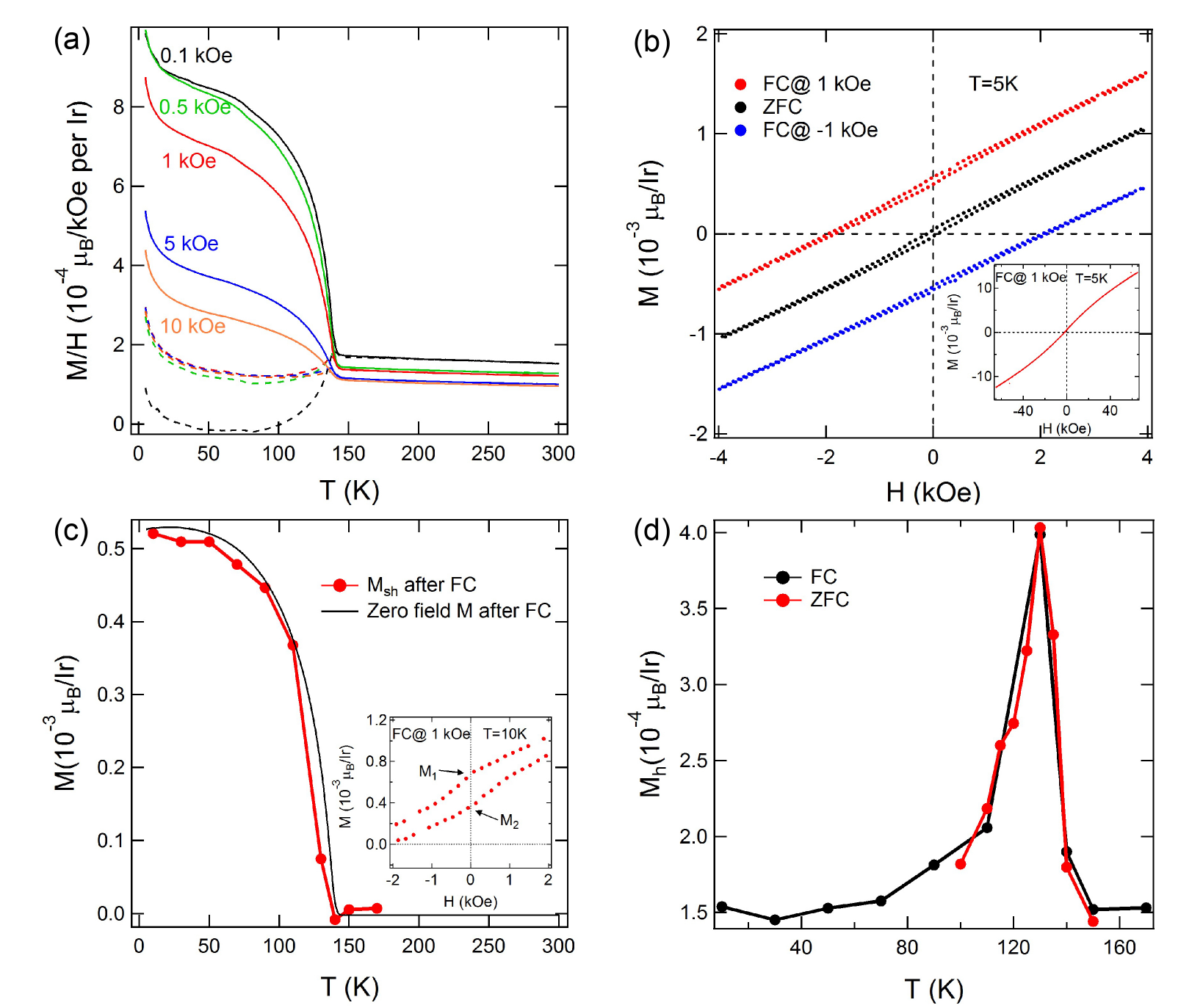}
   	\caption{(color online) Magnetic properties of Lu$_2$Ir$_2$O$_7$: (a) Zero-field cooled (dashed curves) and field-cooled (solid curves) magnetic susceptibilities as a function of temperature at different magnetic fields; (b) Magnetization versus magnetic field (M-H) at 5 K after the sample was cooled down in 1 kOe, 0 kOe and -1 kOe. The inset is the 1 kOe M-H loop in a full sweeping field range. (c) Comparison of the vertical shift $M_{\rm sh}$ and the magnetization measured in zero field after field cooling at 1 kOe. The inset shows an FC M-H loop at 10 K (data taken on different polycrystalline powder from the same batch as (b)). (d) ZFC and FC $M_h$ as a function of temperature.}\label{Fig1}
   \end{figure}

Temperature dependent dc magnetic susceptibility of Lu$_2$Ir$_2$O$_7$ suggests a magnetic transition at T$_N$ $\sim$ 135 K, below which the zero-field cooled (ZFC) and field-cooled (FC) susceptibilities exhibit a clear difference [Fig. \ref{Fig1}(a)]. A similar behavior was reported in other pyrochlore iridates of intermediate/strong electron interactions (e.g. Y$_2$Ir$_2$O$_7$~\cite{exp_method1,muSR1,Iridate227Zhu} and Eu$_2$Ir$_2$O$_7$~\cite{Xray1}) and the magnetic transition was attributed to the onset of long-range AIAO magnetic order.~\cite{muSR2,Xray1,Nakatsuji,anisotropy,Werner} Magnetic-field dependence of magnetization was measured at 5 K, after the sample was cooled down in zero field (i.e. ZFC), in a magnetic field of +1 k Oe and -1 k Oe (i.e. FC). A small magnetic hysteresis loop was observed in all three cases [Fig. \ref{Fig1}(b)], indicating a weak ferromagnetism coexisting with the AFM background. Unlike the ZFC loop, the FC loops show a shift along the magnetization axis with its sign being determined by the polarity of the magnetic field, i.e. +1 kOe gives rise to a positive shift while -1 kOe leads to a negative shift. A cooling-field induced shift of magnetic hysteresis loop was observed earlier in polycrystalline Y$_2$Ir$_2$O$_7$~\cite{Iridate227Zhu} and Sm$_2$Ir$_2$O$_7$~\cite{JPCM} as well. It is also noted that the Hall resistivity versus magnetic field of single crystalline Eu$_2$Ir$_2$O$_7$ thin films also exhibits a vertical shift after the films are cooled down in a magnetic field.~\cite{SciRepJpn} Since the anomalous component of the Hall resistivity in a magnetic material is generally proportional to its magnetization, a shift of magnetic hysteresis loop along the magnetization axis is anticipated in Eu$_2$Ir$_2$O$_7$. Therefore, we argue that such a cooling-field induced shift of hysteresis loop may be a common phenomenon in magnetic pyrochlore iridates.
  
To understand the origin of the shift, we measured magnetization versus magnetic field at different temperatures (10 - 170 K) after the sample was cooled down in a magnetic field of 1 kOe. The shift of hysteresis loop in its magnetization is quantitatively defined as $M_{\rm sh}\equiv \frac{M_1+M_2}{2}$, where $M_1$ and $M_2$ are the two magnetization values at zero field in a hysteresis loop [inset of Fig. \ref{Fig1}(c)]. The $M_{\rm sh}$ decreases monotonically upon warming and reaches zero around $T_N$ [Fig. \ref{Fig1}(c)].  To illustrate the relation between $M_{\rm sh}$ and the magnetic moments that are pinned as a result of field cooling, we carried out another temperature dependence of magnetization measurement in the following protocol: the sample was first cooled down in a field of 1 kOe; then the magnetic field was set to zero using an oscillating mode in the MPMS; finally the magnetization was measured in zero magnetic field upon warming. In such a process, the measured magnetization reveals the magnetic moments that are pinned after the cooling process. As shown in Fig. \ref{Fig1}(c), the magnetization agrees well with $M_{\rm sh}$ at all temperatures, indicating that the vertical shift is clearly due to the pinned magnetic moments. Another parameter $M_{\rm h}\equiv\frac{M_1-M_2}{2}$ is defined to characterize the weak ferromagnetic component. As shown in Fig. \ref{Fig1}(d), the $M_{\rm h}$ has a non-monotonic dependence on the temperature: it increases first and then decreases, yielding a peak around $T_N$. Magnetic hysteresis loops were also recorded after zero field cooling and the corresponding $M_{\rm h}$ shows nearly the same trend as the FC data. It is worth noting that the $M_{\rm h}$ shows some variations among different polycrystalline powder from the same batch, which likely results from stoichiometric inhomogeneity. However, the fact that both the $M_{\rm sh}$ and $M_{\rm h}$ show strong temperature dependence around $T_N$ confirms that the vertical shift of hysteresis loops is an intrinsic feature of the pyrochlore phase, in contrast to the paramagnetic or diamagnetic nature of the impurity phases (i.e. Lu$_2$O$_3$, IrO$_2$, and Ir).~\cite{impurity1,impurity2,impurity3} We note that the enhancement of magnetic hysteresis loop (or coercive field) around $T_N$ was reported in exchange-bias systems in which FM domain/layer is pinned by AFM.~\cite{Leighton, Scholten} The maximum $M_{\rm h}$ is hence an indication of interfacial coupling between the FM droplets and the AFM background which will be discussed in details below.

The weak ferromagnetic component observed earlier in Y$_2$Ir$_2$O$_7$ was attributed to the presence of Ir$^{5+}$ as a result of non-stoichiometry.~\cite{Iridate227Zhu} We carried out X-ray photoelectron spectroscopy to study the oxidation state of Ir in Lu$_2$Ir$_2$O$_7$. As shown in Supplemental Material Figure S2, the iridium spectrum has three components which are attributed to: Ir$^{4+}$, Ir$^{3+}$, and Ir$^0$. The Ir$^0$ is consistent with the presence of iridium metal impurity, resulting from the decomposition of some IrO$_2$ in the solid state reaction process. The Ir$^{3+}$ which may arise from non-stoichiometry of the sample has an electronic configuration of 5d$^6$. The six 5d electrons completely fill the $t_{2g}$ states (or the $J_{\rm eff} = 1/2$ and $3/2$ states), leading to a nonmagnetic state. We show below that the presence of non-magnetic defects in the network of Ir tetrahedra gives rise to ferromagnetic (FM) droplets and the exchange coupling between the FM droplets and the AFM background yields a vertical shift of magnetic hysteresis loop.     

The Hamiltonian for the magnetic moments in a standard magnetic system with strong SOC has the following form,  
\be
	H = \sum_{<i,j>} \left[ J \vec S_i\cdot\vec S_j + \vec D_{ij}\cdot\vec S_i\times \vec S_j + \vec S_i \Gamma \vec S_j \right]\:.\label{JDM}
\ee The first term with $J>0$ represents the antiferromagnetic Heisenberg exchange coupling between the nearest magnetic moments ($\vec S_i$), while the second term is the antisymmetric DM interaction. With the convention for the site indices in Ref.~\onlinecite{DM_AIAO}, the direction of $\vec D$ determines whether the interaction is ``direct"  ($D>0$) or ``indirect" ($D<0$). The third term stands for the anisotropic coupling with a symmetric traceless matrix $\Gamma$. In pyrochlore iridates, the DM interaction is comparable with $J$, while the anisotropic term is much smaller~\cite{Pesin1} and is hence neglected in our calculations. The angular momentum $\vec S$ is treated classically. 

\begin{figure}
	\input{epsf}
	\includegraphics[width=0.8\textwidth]{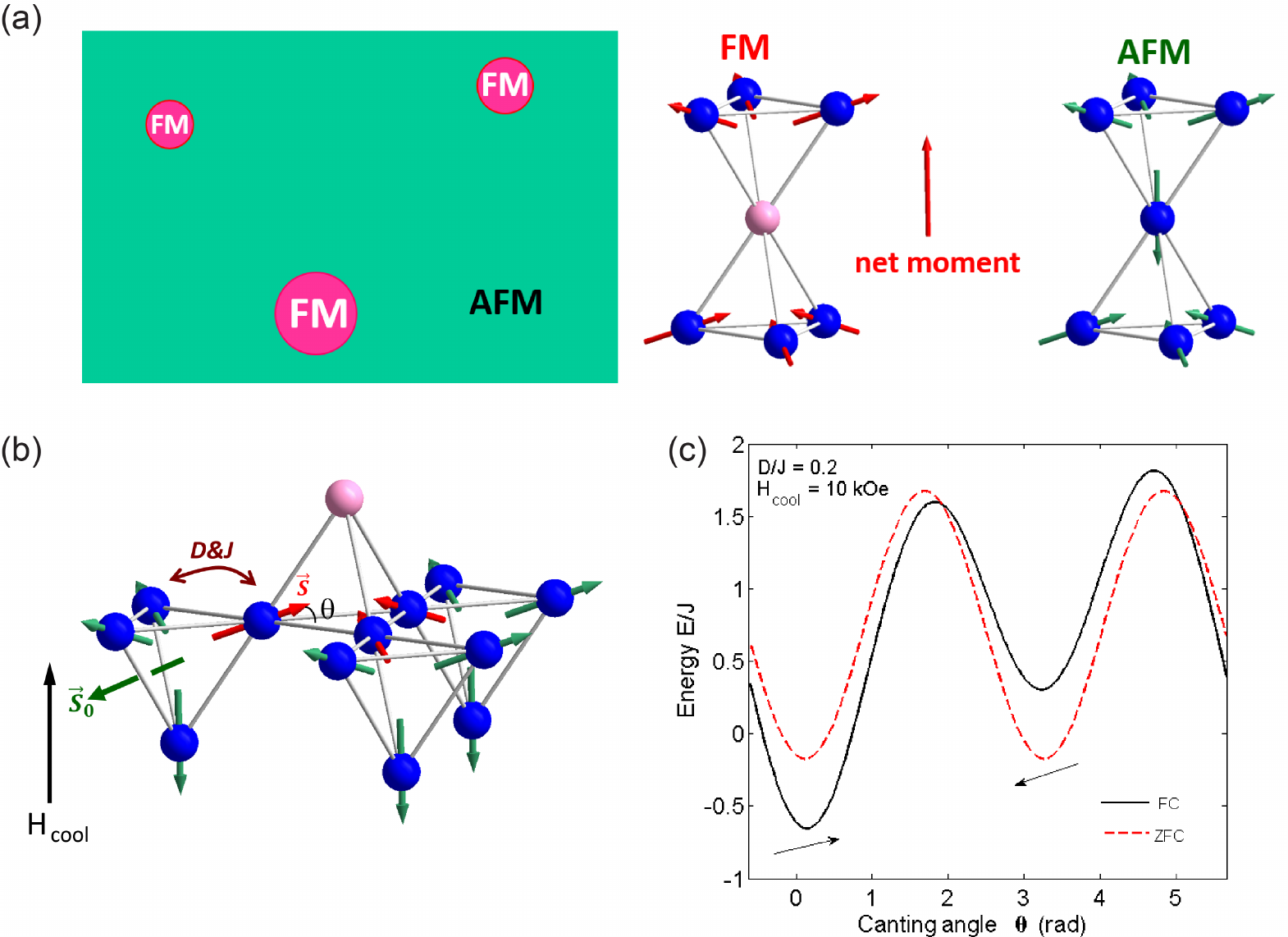}
	\caption{(color online) (a) A schematic picture illustrates the presence of FM droplets in the AFM background: an FM (or canted-AFM) droplet is formed around a non-magnetic Ir defect (pink sphere), while the AIAO magnetic structures are formed in the remaining region. (b) Illustration of magnetic pinning after field cooling: the magnetic moments in an FM droplet (top tetrahedron) are aligned with the cooling field; the remaining moments form AIAO or AOAI order, depending on the orientation of the FM moments. In the meanwhile, the FM moments are pinned by an exchange field due to the Heisenberg and DM interactions from the moments in the AFM background. $\theta$ is the canting angle of the FM moments. The effective moment $\vec S_0$ (green) represents a sum of the three moments in the lower left tetrahedron. (c) energy $E/J$ as a function of $\theta$ in the case of FC (solid) and ZFC (red dashed). The arrows indicate the direction of the moment $\vec S$ in (b).}\label{Fig2}
\end{figure}

In a perfect pyrochlore iridate with intermediate or strong electron interactions (e.g. stoichiometric Y$_2$Ir$_2$O$_7$), the {\it direct} DM interaction stablizes a long-range antiferromagnetic order with an AIAO magnetic configuration.~\cite{DM_AIAO} In the presence of a nonmagnetic defect, the six Ir$^{4+}$ moments surrounding the defect can be considered as an FM (or canted AFM) droplet [Fig. \ref{Fig2}(a)]. Each droplet shall correspond to a magnetic moment of 1 $\mu_{\rm B}$ if the remaining spins are not re-oriented. However, the presence of non-magnetic defect influences the orientations of the remaining spins. Namely, a coplanar configuration (i.e. $\theta=0$) is expected in the limit $D/J\rightarrow 0$, whereas a configuration with $\sin\theta > \frac{1}{3}$ shall occur in the large $D/J$ limit (details will be elaborated in next paragraph). In a field cooling process, the Ir$^{4+}$ moments in a FM droplet [denoted by red arrows in Fig. \ref{Fig2}(b)] are aligned by the cooling-field $H_{\rm cool}$ via Zeeman coupling. The remaining Ir$^{4+}$ moments form either AIAO or its time-reversal counterpart AOAI order [denoted by green arrows in Fig. \ref{Fig2}(b)] and their orientations are pre-determined by the moments in the FM droplet. On the other hand, the Heisenberg and the DM interactions between the Ir$^{4+}$ moments at the interface of the FM droplets and the AFM background give rise to an additional exchange field that acts on the FM moments. This exchange field pins the FM moments, giving rise to an exchange-bias like behavior. 

To quantitatively study the dependence of $M_{\rm sh}$ on cooling-field $H_{\rm cool}$, we model the magnetic energy $E$ of an FM droplet [i.e. the top tetrahedron in Fig. \ref{Fig2}(b)] as a function of canting angle $\theta$ (i.e. the angle made by the spin $\vec S$ with the bottom face of the same tetrahedron). In the absence of non-magnetic defect, the three canted moments at the bottom face collectively balance the moment at the top vertex, so the canting angle has $\sin\theta = 1/3$. Because of the three-fold rotational symmetry, it suffices to consider the lower-left moment $\vec S$ as a representative for the three moments. Based on Eq.~\ref{JDM}, we have obtained the following expression in which all terms irrelevant to dynamics are neglected, 
\be
	E = \frac{3}{2} J\sin^2\theta -\frac{\sqrt{3}}{2} D \sin(2\theta+\theta_1) - Jf(H_{\rm cool})\cos(\theta-\theta_0)\label{E_tot}\:.
\ee The first term proportional to $J$ results from rewriting the sum of inner products $J\sum\vec S_i\cdot\vec S_j = (1/2)J|\sum\vec S|^2$. This term favours a coplannar configuration corresponding to $\sin\theta = 0$. The second term along with the constant angle $\sin\theta_1 = 1/\sqrt{3}$ results from employing the explicit expressions for the normalized $\vec D$ vectors specified in Ref.~\onlinecite{DM_AIAO}. This term {\it alone} gives rise to a canting angle of $\sin\theta\approx 0.46$ in the FM droplet, larger than the canting angle with $\sin\theta_0 = 1/3$ in a perfect tetrahedron. The last term stands for the antiferromagnetic coupling between $\vec S$ and the rest three moments (denoted by green arrows) in the lower-left tetrahedron. We denote the sum of those three moments by $\vec S_0$. To simplify the calculation, we adopt the mean-field picture: given the direction of $H_{\rm cool}$ shown in Fig. \ref{Fig2} (b), the ``averaged" magnitude $|S_0|$ monotonically increases from zero to a saturated value as $H_{\rm cool}$ is increased from zero (ZFC) to some critical value $H_0$. Thus, it suffices to consider the effect of cooling-field through the fitting functional,~\cite{SciRepJpn,PRB05}
\be
	|\vec S_0| := f(H) = c_1{\rm tanh}(\frac{H}{H_0}) + c_2H\:,
\ee where $c_1$ and $c_2$ are positive parameters to be determined by matching with measurement data. Qualitatively, $H_0$ is the critical field beyond which the FM droplets are aligned and the ``averaged" $S_0$ stops increasing rapidly, while its value is to be determined by data, too. In the absence of $H_{\rm cool}$, i.e. ZFC, the third term in Eq.~\ref{E_tot} vanishes and $E_{\rm tot}$ has two identical local minima corresponding to two $\theta$ values that are separate by $\pi$ [dashed curve in Fig. \ref{Fig2}(c)]. This can be seen from the symmetry of the first two terms in Eq.~\ref{E_tot} under $\theta\rightarrow\theta+\pi$. As such, the moments in the FM droplet have an equal chance to point at one direction or at its opposite, and such situation is independent of the ratio $D/J$. On the other hand, in the presence of $H_{\rm cool}$ (i.e. FC), the energy has two different local minima ($E_p$ and $E_{ap}$) that are separated by an energy barrier (solid curve in Fig. \ref{Fig2}(c)). The difference between $E_p$ and $E_{ap}$ is significant in determining the magnetization in the sweeping process. Following the discussion in Ref.~\onlinecite{PRB05}, when the sweeping field changes from a very large and positive (negative) value to zero, the FM droplets initially occupying the state of $E_p$ ($E_{ap}$) will partially switch to the state of $E_{ap}$ ($E_p$) to reach an equilibrium state. Although the details depend on the tunneling mechanism and the barrier between two states, the average of the two intersections, $M_1$ and $M_2$, from the two paths of varying the $H$ only depends on the energy difference between $E_p$ and $E_{ap}$. Namely, we can write ~\cite{PRB05}   
\be
	M_{\rm sh} \propto \frac{\sin\theta_p\ e^{-\beta E_p} + \sin\theta_{ap}\ e^{-\beta E_{ap}}}{e^{-\beta E_p} + e^{-\beta E_{ap}}}\:,
\ee with $\theta_{p(ap)}$ representing the canting angle corresponding to the lower (higher) energy minimum. $\beta = k_BT_0$ is associated with the measurement temperature $T_0$.

\begin{figure}
\input{epsf}
\includegraphics[width=0.5\textwidth]{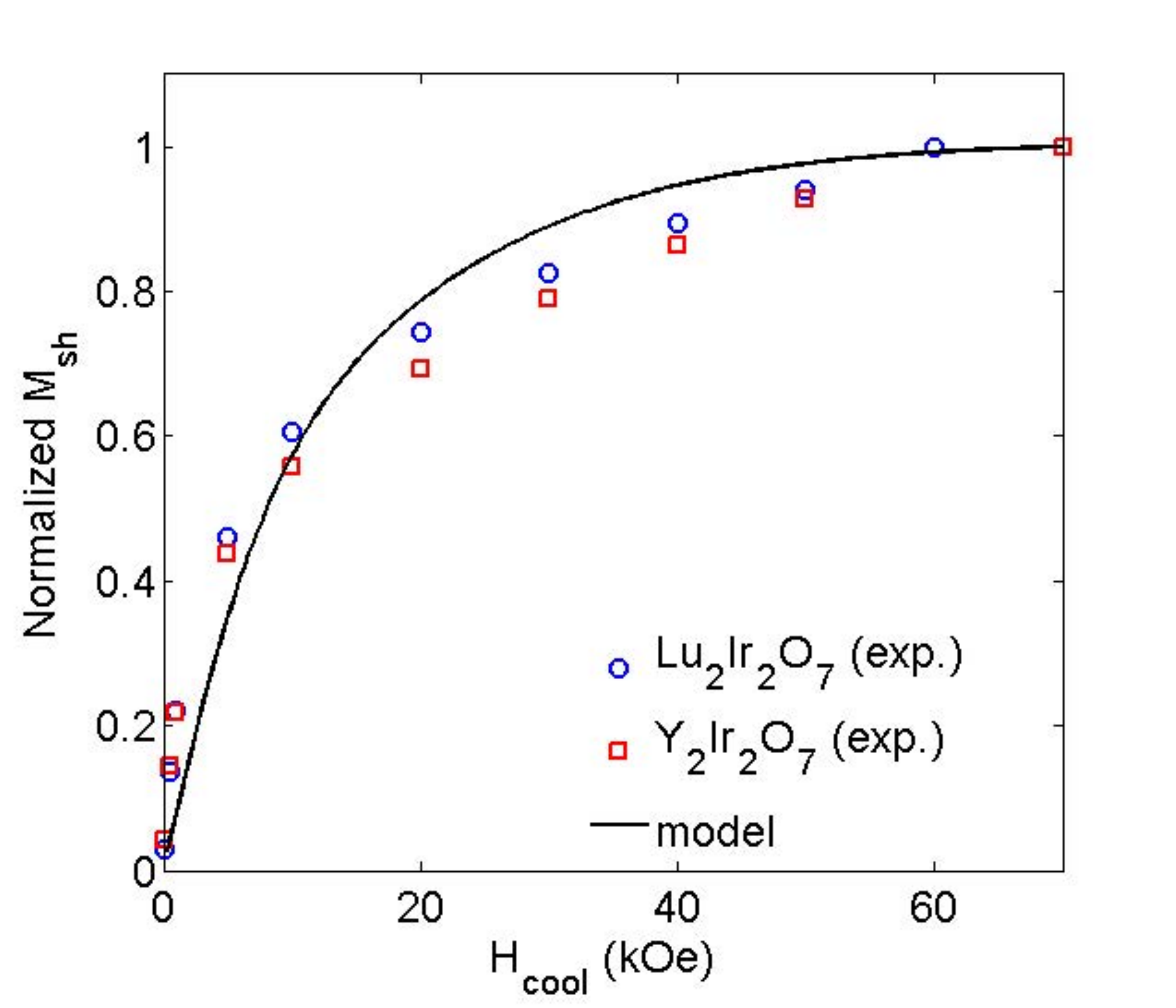}
\caption{Normalized $M_{\rm sh}$ against the cooling-field $H_{\rm cool}$ for Lu$_2$Ir$_2$O$_7$ (blue circle) and Y$_2$Ir$_2$O$_7$ (red square). The solid line represents the calculated $M_{\rm sh}$ with parameters of $H_0 =$ 9 kOe, $c_1 =$ 0.3, $c_2 =$ 0.025, $k_{\rm B}T_0$ = 0.75 $J$, and $D/J$=0.2.}\label{Fig3}
\end{figure}

We computed $M_{\rm sh}$ vs $H_{\rm cool}$ curves with different parameters in Supplemental Material Figure S3. While the $M_{\rm sh}$ is dependent on other parameters (mostly on $k_{\rm B}T_0/J$), it is nearly insensitive to the $D$ value. This behavior can be understood based on Eq.~\ref{E_tot}. In brief, the magnetic energy of the FM droplet always has a pair of degenerate local minima and their canting angles differ by 180 degrees when the third term is neglected (i.e. in the case of ZFC). Thus, different values in $D$ only shift the canting angles corresponding to the local minima but do not affect the energy difference between the two minima when $H_{\rm cool}$ is present. Figure~\ref{Fig3} shows a good agreement between the measured $M_{\rm sh}$ and the computed data with parameters of $c_1 =$ 0.3, $H_0 =$ 9 kOe (or 0.9 T), $c_2 =$ 0.025, $k_{\rm B}T_0$ = 0.75 $J$ and $D/J$=0.2 (the least sensitive parameter). The coefficient $c_1$ = 0.3 reflects roughly the polycrystalline nature of the sample in which the crystals orient randomly with respect to $H_{\rm cool}$. The $H_0 = 0.9$ T is close to the critical field ($\approx$ 1 T) observed in the measurement of linear magnetoresistance coefficient $\alpha$ for Eu$_2$Ir$_2$O$_7$ where $\alpha$ reaches $tanh(1)\approx$0.76 of its saturated value at the field of 1 T.~\cite{SciRepJpn} The $c_2$ = 0.025 is indeed small so $M_{\rm sh}$ may saturate at a large cooling-field. The ratio of temperature to antiferromagnetic coupling $k_{\rm B}T_0$ = 0.75 $J$ corresponds to $J\approx$ 1.1 meV at a measurement temperature of 10 K. This value is lower than the exchange coupling in Sm$_2$Ir$_2$O$_7$ (~27 meV) that was determined by resonant magnetic x-ray scattering.~\cite{SmIr}. The $J$ and $D$ values in pyrochlore iridates, according to Ref.~\onlinecite{DM_ABINITO}, are very sensitive to the Ir-O-Ir bonding angle, so the magnetic interaction values vary among the pyrochlore iridate family. It is worth mentioning that while magnetic domain walls are expected in pyrochlore iridates, the $M_{\rm sh}$ is unlikely to be contributed from defect pinning of the domain walls. Indeed, increasing cooling field tends to reduce the number of domains \cite{Science2015} and hence decreases $M_{\rm sh}$ if the domain wall pinning picture holds, which contradicts with the monotonic increase of $M_{\rm sh}$ shown in Figure 3.

\begin{figure}
\input{epsf}
\includegraphics[width=0.5\textwidth]{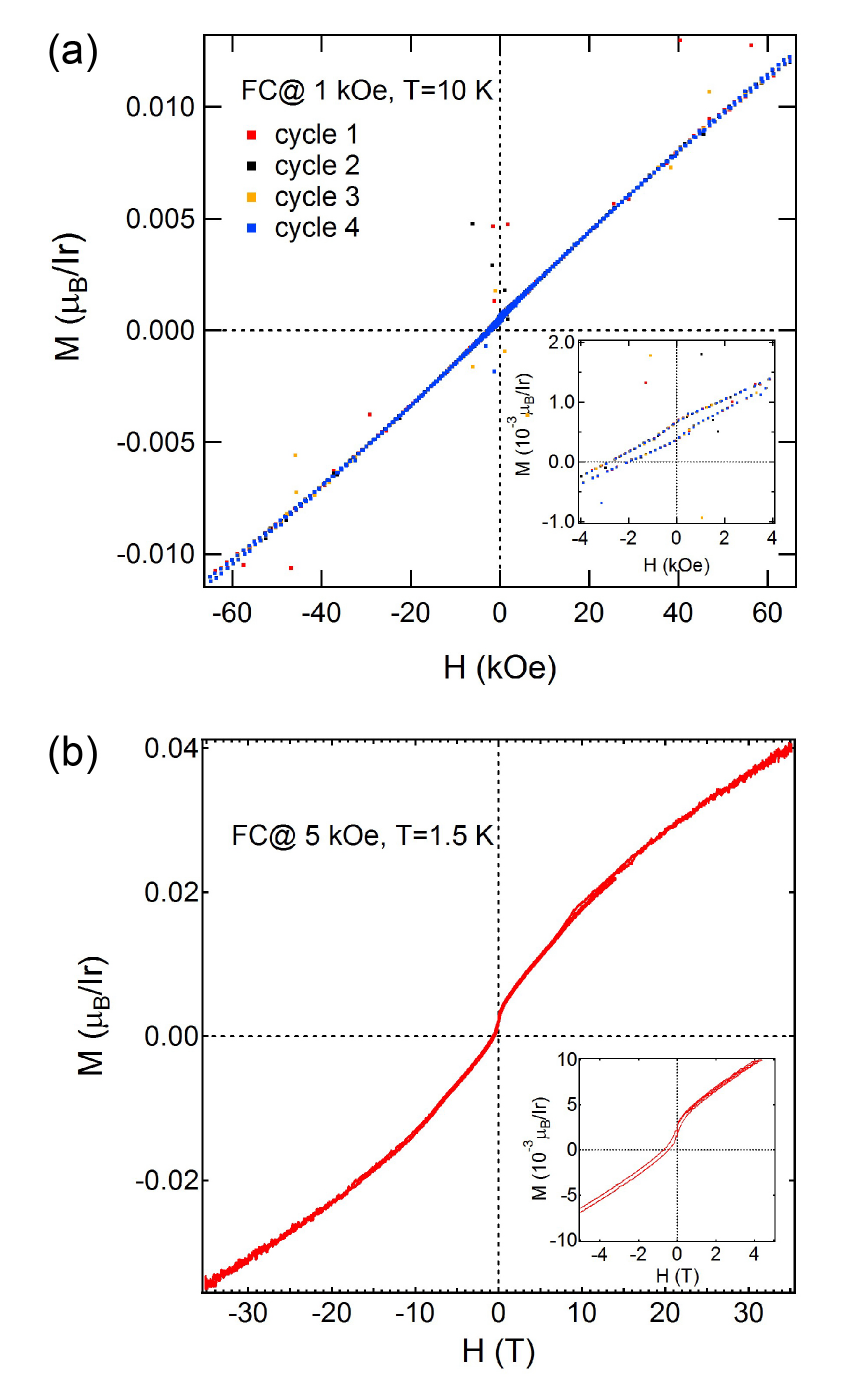}
\caption{(color online) (a) M-H loops of Lu$_2$Ir$_2$O$_7$ measured at 10 K in a consecutive manner, after 1 kOe field cooling. The inset shows the hysteresis loops at low fields. The scattered data points should be due to the magnetization measurement itself instead of being an intrinsic property of the sample. (b) A high-field M-H curve of Y$_2$Ir$_2$O$_7$ measured at 1.5 K after 5 kOe field cooling. The low field data are presented in the inset.}\label{Fig4}
\end{figure}

Lastly, we note that the magnetic hysteresis loops in a conventional exchange-bias system are horizontally shifted.~\cite{EB1, EB2} The vertical shift observed here suggests that the magnetic pinning is very robust in the pyrochlore iridate system. We demonstrate next the robustness of magnetic pinning against sweeping cycle and sweeping field. Figure \ref{Fig4}(a) presents four magnetic hysteresis loops that were measured in a consecutive manner, after the sample was cooled down to 10 K in a magnetic field of 1 kOe. The M-H loops overlap with each other and no training effect is observed, indicating the pinning of magnetic moment is indeed stable. We further carried out a high field measurement on a Y$_2$Ir$_2$O$_7$ sample up to 35 T. As shown in Figure \ref{Fig4}(b), the vertical shift of magnetic hysteresis loop persists even after the sample was subject to a field swept from 35 to -35 T. The robustness of magnetic pinning may be related to the octupolar nature of the AIAO order. Indeed, both the dipole and quadrapole terms vanish in an AIAO state and the resulted octupolar term has a susceptibility that is described by a third-rank tensor.~\cite{Arima,NatPhysFu} The third-order tensor coupling with the magnetic field is presumably weaker than the exchange couplings $J$ and $D$ between dipole moments, which is possibly why the pinned moments are not switchable by an external magnetic field. 

\section{CONCLUSIONS}
In summary, we report on a cooling-field induced vertical shift of magnetic hysteresis loop and its possible origin in pyrochlore iridates that contain non-magnetic Ir defects. In the presence of nonmagnetic defects, FM droplets are formed out of the background AIAO/AOAI spin ordering. The vertical shift can be understood quantitatively based on an exchange-bias like model in which the moments at the shell of the FM droplets are pinned by the AIAO AFM background. The exchange bias in the magnetization loop is stable and robust against sweeping cycle and sweeping field, which is possibly associated with the magnetic octupolar nature of the AIAO order.

\section*{ACKNOWLEDGMENTS}
We thank Professors L. Hozoi, W. Tong, and G. Chen for helpful discussions, and Professors L. Pi and J. Long for experimental assistance. S.X.Z. acknowledges Indiana University (IU) College of Arts and Sciences for startup support. C.K.L. was supported by the Taiwan Ministry of Science and Technology through Grant No. 03-2112-M-003-012-MY3. H.D.Z. thanks the support from NSF-DMR through Award DMR-1350002. The work at NHMFL is supported by NSF-DMR-1157490 and  the State of Florida. We are grateful to the Indiana University Molecular Structure Center for the use of XRD facility (NSF CHE-1048613) and  the Nanoscale Characterization Facility for XPS instrument (NSF DMR 1126394). We also thank the High Magnetic Field Laboratory, Chinese Academy of Sciences for partial use of MPMS. 


\end{document}